# POLARIMETRIC METHOD OF PLASMA DIAGNOSTICS

**Koryun B. Oganesyan**[1,*], **Krzysztof Dzierżęga**[2] **and Peter Kopcansky**[3]

[1] A. Alikhanyan National Lab, Yerevan Physics Institute, Yerevan, Armenia

[2] Marian Smoluchowski Institute of Physics, Jagiellonian University, Kraków, Poland

[3] Institute of Experimental Physics SAS, Kosice, Slovakia

*bsk@yerphi.am

The polarization plane rotation angle of probe signal in plasma is calculated. The estimates of the residual gas average density and the average magnitude of the magnetic field in a cesium plasma based on the effects of Faraday and Cotton-Mouton in probe laser field has been found. It is shown, that polarization plane rotation angle depends on resonance detuning of incident laser wave and transition frequency of medium atoms, magnetic field strength, and matrix element.

## 1. INTRODUCTION

One of the main tasks set before plasma researchers is the problem of using thermonuclear energy [1]. At the first stages of development, the main attention was paid to the determination of such plasma parameters as the concentration of charged particles and (electrons and ions ), their temperatures and , the degree of ionization, etc. Now more and more attention of researchers is attracted by the problem of determining the residual density of the neutral component of the plasma and the magnitude of the magnetic field. Of particular importance is the determination of these parameters in tokamaks [2].

Plasma diagnostic methods can be classified according to different criteria. For example, depending on whether the sensitive sensor is in contact with the medium under study or not, it is customary to call measurement methods contact or non-contact. In the modern plasma experiment, there is a clear desire for non-contact diagnostics of plasma. In this case, both passive diagnostics are developed - the electromagnetic field or corpuscular radiation emitted by the object under study is analyzed, and active diagnostics - the probing beam of radiation or particles scattered by the plasma is analyzed [3-6].

Magnetic fields in tokamak plasma facilities are created either by currents in the plasma itself (poloidal field) or by currents in external conductors (toroidal field).

To measure the magnetic field from the Faraday rotation of the plane of polarization on conduction electrons, a sufficiently dense plasma is required. Indeed, the angle of rotation per unit length ( z =1cm ) can be estimated from the formula

$$\varphi^0 = 1.5 \cdot 10^{-20} \lambda^2 B N_e$$

where $\lambda$ is the radiation wavelength in microns, B is the magnetic induction in KGs, $N_e$ is the electron density.

In the experiment, the angles of rotation $\varphi = 10^{-1}$ degrees can be registered. With a magnetic field B = 30 KGs and a wavelength of optical radiation $\lambda = 10^{-1}$ μm, such a rotation requires density $N_e \sim 10^{19} \div 10^{20} at/cm^3$. Since the angle of rotation is proportional to the square of the wavelength, it is preferable to use not optical, but microwave radiation. In addition, to measure the magnitude of the magnetic field by this method, it is necessary to know the independently measured electron density.

The method of Zeeman broadening of the spectral lines emitted by a plasma also has its drawbacks. In high-temperature plasmas, the Zeeman broadening is often overlapped by the Doppler or Stark mechanisms of line broadening. This circumstance plays a particularly important role for a plasma consisting of light atoms. Indeed, for a hydrogen plasma, at temperatures of $T \sim 100$ eV and a magnetic field B = 30 KGs, the Doppler broadening is of the order of $\gamma_D \sim 10 \, cm^{-1}$, while the Zeeman broadening is of the order of $\gamma_Z \sim 1 \, cm^{-1}$. Note, that both of the above methods make it possible to determine the integral value of the magnetic field.

Under normal conditions, the gaseous medium is an optically isotropic medium. Optical anisotropy in a gas can arise when the gas is placed in a magnetic field (Faraday and Cotton-Mouton effects), under the action of intense laser radiation of elliptical or circular polarization. Let us consider the last phenomenon.

Intense laser radiation passing through a gaseous medium. changes as the absorption coefficient, and the refractive index of the medium. If a wave has a degree of circular polarization other than zero, then the changes in the absorption coefficient and refractive index caused by it are different for different circular polarizations of light. The first circumstance leads to the induced circular dichroism of the medium, the second is to induced birefringence. If a test linearly polarized signal is passed through such a medium, then its polarization will change at the exit from the medium; due to dichroism, plane-polarized light will turn into elliptically polarized, due to birefringence, a rotation of the plane occurs. The above phenomena in a gas have a pronounced resonant character and become actually observable under the condition that the frequency of laser radiation is close to the frequency of the atomic transition.

In this article, we propose a method for determining the magnitude of the magnetic field in tokamaks - the polarization method of resonant rotation of the polarization plane on neutral plasma atoms. In the presence of a magnetic field, the polarization of the probe signal changes due to the Faraday and Cotton-Mouton effects. These effects are considered below.

## 2. DIAGNOSTICS OF THE AVERAGE MAGNETIC FIELD

The value of the magnetic field in plasma device can be obtained on the basis of the magnetic rotation of the plane of polarization of the probing signal. If a light beam propagates along the direction of magnetic field lines, then a Faraday rotation of the polarization plane occurs, if the directions of propagation of the light wave and the magnetic field are mutually perpendicular, then the polarization of the wave changes due to the Cotton-Mouton effect. Let us consider these effects separately.

### a) FARADAY EFFECT

We assume that the light beam propagates axially along the direction of the magnetic field lines. In a magnetic field, due to the Zeeman effect, the splitting and shift of the energy levels of the cesium atom occurs. Figure 1 shows the example of Zeeman splitting and shift for the resonant transition $6S_{1/2} \to 6P_{1/2}$ ( $\lambda = 8943{,}50 A°$ )

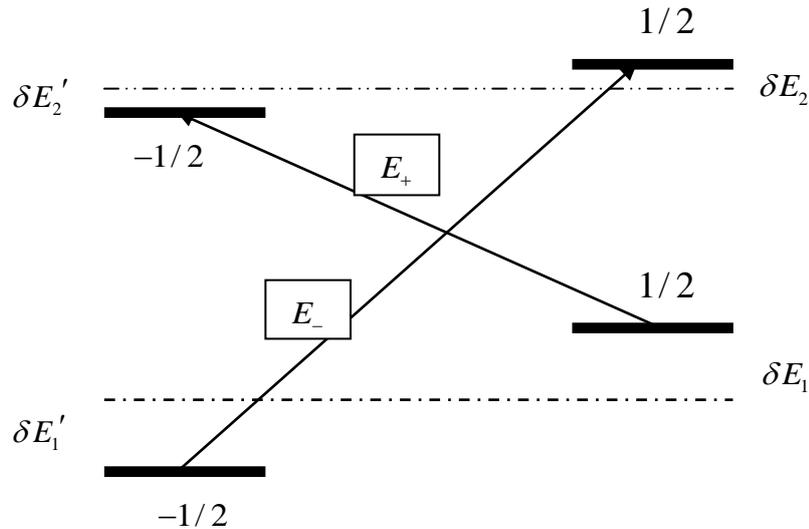

Fig.1. Resonance splitting and shift of energy levels of Cesium atoms (resonance transition $6S_{1/2} \to 6P_{1/2}$ ). The dotted line indicates the position of the levels in the absence of a magnetic field. Energy shifts

$$\delta E_1 = -\delta E_1' = \mu_0 H, \qquad \delta E_2 = -\delta E_2' = \frac{\mu_0 H}{3}$$

A light wave propagates along the z axis and is polarized in the xy plane. It is convenient to represent its polarization in the form of two circular components $E_\pm = E_x \pm i E_y$. The right circular component of the wave $E_-$ interacts with sublevels $M_1 = -1/2 \to M_2 = 1/2$; the left component $E_+$ - with sublevels $M_1 = 1/2 \to M_2 = -1/2$.

The shortened Maxwell equations for the circular components of the wave will have the form

$$\frac{dE_+}{dz} = 2iq \frac{\varepsilon}{\varepsilon - \frac{4}{3}\mu_0 H} E_+$$

$$\frac{dE_-}{dz} = 2iq \frac{\varepsilon}{\varepsilon + \frac{4}{3}\mu_0 H} E_- \qquad (1)$$

$$q = \frac{\pi N \omega |d|^2}{6c\hbar(\Delta \nu)}$$

where N – density of atoms, $\omega$ - wave frequency, $d$ - matrix element, $\varepsilon = \omega - \omega_0$ - resonance detuning, $\omega_0$ is atomic transition frequency and $\omega$ is wave frequency.

It can be seen from the equations that the refractive indices for the right and left circular components of the wave differ from each other. This difference leads to a rotation of the plane of polarization of the probe signal $\varphi = \frac{\omega}{c} \frac{n_+ - n_-}{2} l$, where $n_\pm$ are refractive indexes of left and right circular components of wave, $l$ length of medium. If, at the entrance to the medium, the wave was polarized along the x axis ($E_+ = E_-$), at the exit its plane will rotate by an angle

$$\varphi = qz \left[ \frac{\varepsilon}{\varepsilon - \frac{4}{3}\mu_0 H} - \frac{\varepsilon}{\varepsilon + \frac{4}{3}\mu_0 H} \right] \qquad (2)$$

If the resonance detuning $\varepsilon$ is large compared to $\mu_0 H$, formula (3.6) takes the form

$$\varphi = \frac{8}{3} \frac{\mu_0 H}{\varepsilon} qz \qquad (3)$$

Substituting thr above expression for q into this formula, we obtain

$$\varphi \approx 2 \cdot 10^{-14} \frac{N}{\Delta \nu} \cdot \frac{\mu_0 H}{\Delta \nu} \cdot z \qquad (4)$$

Knowing the density of atoms, it is possible to determine the magnitude of the magnetic field, or rather its average characteristic $\int H dz$, by the magnitude of the angle of rotation.

Substituting the value of the Bohr magneton, formula (4) can be rewritten as

$$\varphi \approx 9.34 \cdot 10^{-19} \frac{N(\text{cm}^{-3})}{(\Delta \nu)^2 (\text{cm}^{-1})} z(\text{cm}) \cdot H(\text{Oe}) \qquad (4^*)$$

It is clear, that the mechanism of light polarization is directly related to the shape of the spectral line of atoms. Below, we will consider in more detail on the various mechanisms of broadening of the spectral lines of neutral atoms in a plasma.

Let us estimate the minimum magnetic field that can be detected in a cesium plasma under the following experimental conditions N= $10^{13} cm^{-3}$, z=10, $\varphi = 10^{-2}$ radians, T=10ev ( $\Delta v_d = 0.4 cm^{-1}$ ), $\Delta v = 0.5 cm^{-1}$. Then for the minimum magnetic field we have $H_{min}$ =30 KOe. Formula (4*) allows us to follow the change in the observed magnetic field in other conditions in an elementary way.

Let us also consider the situation in tokamaks, when a cesium impurity is also introduced into the plasma to diagnose the average poloidal magnetic field in a hydrogen plasma. Let $H_p$ =3KOe, T=100ev, ( $\Delta v_d = 14 cm^{-1}$ ), z=10cm, $\varphi = 10^{-2}$. Then the required density of neutral cesium atoms is of the order N=$10^{13} cm^{-3}$ .

### b) COTTON - MOUTON EFFECT

Let us assume that the directions of propagation of the light wave and the magnetic field are mutually perpendicular. The z axis is directed along the direction of the magnetic field, and the y axis is directed along the direction of wave propagation. The wave can be polarized in the xz plane. It is convenient to decompose the electric field strength vector into two linear components; $E_z$ – polarized along the z axis, interacts with magnetic sublevels $\Delta M = 0$; the component $E_x = \frac{E_+ + E_-}{2}$ is polarized along the z axis with magnetic sublevels $\Delta M = \pm 1$ (Fig. 1).

Shortened Maxwell equations for linear components $E_z$ and $E_x$ will have the form

$$\frac{dE_z}{dy} = 2iqE_z \left[ \frac{\varepsilon}{\varepsilon - \frac{2}{3}\mu_0 H} + \frac{\varepsilon}{\varepsilon + \frac{2}{3}\mu_0 H} \right]$$

$$\frac{dE_x}{dy} = 2iqE_x \left[ \frac{\varepsilon}{\varepsilon - \frac{4}{3}\mu_0 H} + \frac{\varepsilon}{\varepsilon + \frac{4}{3}\mu_0 H} \right]$$

(5)

It can be seen from the equations obtained that the gaseous medium in a magnetic field perpendicular to the direction of light wave propagation behaves like a uniaxial crystal. A wave polarized at the input along the z (or x) axis does not change its linear polarization. Any other linear polarization changes as it passes.

Let us assume that at the input the wave is polarized at an angle of 45 degree to the z axis. The angle of rotation of the polarization plane at the output will be equal to

$$\varphi = 2qz \left[ \frac{\varepsilon^2}{\varepsilon^2 - \left(\frac{4}{3}\mu_0 H\right)^2} - \frac{\varepsilon^2}{\varepsilon^2 - \left(\frac{2}{3}\mu_0 H\right)^2} \right] \qquad (6)$$

If the resonance detuning $\varepsilon$ is large compared to $\mu_0 H$, then formula (6) takes the form

$$\varphi = \frac{8}{3}\left(\frac{\mu_0 H}{\varepsilon}\right)^2 qz \qquad (7)$$

Comparing this formula with formula (4), we see that the Cotton-Mouton effect is proportional to the square of the magnetic field, while the Faraday effect depends linearly on the magnitude of the magnetic field. Substituting into (7) the value of q from (1) and the value of the Bohr magneton:

$$\varphi \approx 1.45 \cdot 10^{-23} \frac{N(\text{cm}^{-3})}{(\Delta\upsilon)^3(\text{cm}^{-1})} z(\text{cm}) \cdot H^2(\text{Oe}) \qquad (7^*)$$

It can be seen from the obtained formula that in order to detect magnetic fields of the order of ten kilooersteds at $N = 10^{12} - 10^{13} cm^{-3}$ z = 10 cm, it is necessary to have relatively low temperatures and heavy elements (for example, cesium), so that the detuning $\Delta\upsilon$ (limited from below by Doppler broadening) is small $\Delta\upsilon \ll 1 cm^{-1}$. This order of detuning for cesium atoms is obtained at T=100 eV.

Note, also that the Faraday effect on electrons at the considered densities ( $N_i \sim N_e \approx 10^{13} \div 10^{14} cm^{-3}$ ) and magnetic fields (H = 30 KOe) is negligible compared to the Faraday effect on neutral resonant atoms.

## CONCLUSIONS

The average magnetic field magnitude and neutral atom average density in cesium plasma are estimated. It is shown the acceptability of proposed method to diagnostics of average density of neutral atoms and average magnitude of magnetic field in tokamaks. In conclusion we note, that the Cotton-Mouton effect is weaker than the Faraday effect and, therefore, in those devices that allow us to direct the probing laser beam along the magnetic field, it is advisable to use the Faraday effect.